\documentclass[prb,aps,showpacs,amsmath,twocolumn,10pt]{revtex4-1}
\usepackage{bookmath} % definitions and shortcuts
\usepackage{graphicx}
\usepackage{color}
\usepackage[caption=false]{subfig}
\newcommand{\cecoin}{CeCoIn$_5$} 
\newcommand{\kbtf}{$\kappa$-(BEDT-TTF)$_2$Cu(NCS)$_2$}
\newcommand{\sign}{\mbox{sgn}}
\newcommand{\vvec}[1]{ \overset{\text{\scriptsize$\leftrightarrow$}}{#1} }
\begin{document}
%~~~~~~~~~~~~~~~~~~~~~~~~~~~~~~~~~~~~~~~~~~~~~~~~~~~~~~~~~~~~~~~~~~~~~~~~~~~~~~~%
\title{Spin susceptibility of Andreev bound states} 
%\title{Spin susceptibility and spin-lattice relaxation rate of a superconducting domain wall}
%\title{Spin Susceptibility and Spin-Lattice Relaxation Rate in non-uniform Superconductors}

\author{B. M. Rosemeyer, Anton B. Vorontsov}
\affiliation{Department of Physics, Montana State University, Montana 59717, USA}

\date{\today}

\begin{abstract}
We calcuate electronic spin susceptibility and spin-lattice relaxation rate 
in singlet superconductor near a pairbreaking surface, or in a domain wall of the order parameter. 
We directly link presence of high-density Andreev bound states in the inhomogeneous region, combined with coherence factors, 
to enhancement of the susceptibility above the normal state's value for certain $\vq$ vectors.  
Beside the dominant peak at ferromagnetic vector $q=0$, we find significant enhancement of 
antiferromagnetic correlations at vectors $q\lesssim 2 k_f$, 
with $\vq$ \emph{along} the domain wall in $S$-wave superconductor, 
and \emph{across} domain wall in $D$-wave (nodes along the wall). 
%\blue{with nodes along the wall, and \emph{along} domain wall when the maximum gap is along the domain wall}.  
These features are destroyed by applying moderate Zeeman field that splits the zero-energy peak. 
We solve Bogoliubov-de Gennes equations in momentum space and 
our results deviate from the lattice models investigated previously. 
%the DW and we use the Andreev approximation to find a phase relation between two bound states connected by a wave vector
%$\vq$, that when satisfied, can lead to enhancement of the susceptibility above the normal metal value. 
Large enhancement of the spin-lattice relaxation rate $T_1^{-1}$ at the domain wall 
provides clear signature of the quasiparticle bound states, 
and is in good agreement with recent experiment in organic superconductor \kbtf. 
%for temperatures and fields that allow transitions between bound states of different spins or bound states and continuum states. 
%We consider singlet superconductors with S or $D$-wave pairing and solve the at various temperature and Zeeman field.  
\end{abstract} 

%\pacs{74.20.Rp,74.25.Ha}

\maketitle

%~~~~~~~~~~~~~~~~~~~~~~~~~~~~~~~~~~~~~
\section{Introduction}
\label{sec:intro}
%~~~~~~~~~~~~~~~~~~~~~~~~~~~~~~~~~~~~~
%

Soon after formulation of the BCS theory\cite{BCS}
Fulde, Ferrel\cite{PhysRev.135.A550} and Larkin, Ovchinnikov\cite{larkin1965inhomogeneous} (FFLO)
pointed out that nonuniform superconducting states play an important role 
in strong magnetic fields or in magnetically-active materials. 
The most characteristic feature of nonuniform superconductors are distinct quasiparticle states 
that lie inside the energy gap of the bulk phase. 
They appear at pairbreaking surfaces in unconventional superconductors,\cite{CHu1994} 
in vortex cores,\cite{CAROLI1964307} 
heterostructures,\cite{MEschrig2015}
and recently they were connected to topological properties of the order parameter.\cite{Tanaka2012,Mizushima2016}
%Recently, there have been proposals to describe properties of the cuprate superconductors 
%with high-momentum pair-density-wave condensates.\cite{PLee2014}
Generally known as Andreev bound states (ABS) they are localized, for example, near 
a surface of a superconductor and decay into the bulk 
within a few coherence lengths $\xi_{c} = \hbar v_f/2\pi k_B T_c$.
If the bound states are all concentrated at one energy, 
producing a strong peak in the density of states (DOS), they dramatically change properties 
of the surface layer. 

One important question is how the bound states affect magnetic properties of a material, in particular 
electronic spin susceptibility $\chi$ and spin-lattice relaxation rate $T_1^{-1}$. 
For example, in triplet superfluid $^3$He these observables may provide a way to probe 
surface Majorana states.\cite{Nagato:2009chi,Chung:2009t1}
In singlet superconductors they may be used to manipulate magnetic properties of 
the surface layer, or help prove or disprove existence of FFLO phases. 
This last goal is particularly relevant for several materials. 
In heavy-fermion superconductor \cecoin, can an FFLO phase be the origin of 
coexistence\cite{cecoin5_Kenzelmann,cecoin5_Kenzelmann2,Gerber2014} of antiferromagnetism (AFM) and superconductivity?
On the other hand, is recently observed\cite{Mayaffre2014} enhancement of relaxation rate $T_1^{-1}$ 
in organic superconductor \kbtf, indeed explained by Andreev bound states at FFLO domain walls?  

Previous investigations of how nonuniform FFLO order parameter (OP) 
structures influence magnetic properties used quasiclassical techniques,
and real-space lattice Hamiltonians. 
The quasiclassical calculations\cite{Burkhardt1994,Vorontsov:2006fc} 
show about 10\% enhancement of uniform 
magnetization inside FFLO domains at high fields where the FFLO phase appears. 
However, this technique cannot say anything about antiferromagnetic correlations 
with ordering vectors beyond $q \approx 1/\xi_c$. 
Several two-dimensional lattice Hamiltonians have been solved via Bogoliubov-de Gennes (BdG) equations 
to investigate co-existence of AFM order and FFLO states.\cite{Yanase2009, Yanase2009abs, Marcin2009, Yanase2011} 
This approach can treat modulations on the order of Fermi momentum $q \sim 2k_f$. 
It was found that incommensurate spin-density wave (SDW) order
can be induced inside the FFLO phase.\cite{Yanase2009,Yanase2009abs} 
Other calculations show that transverse and longitudinal susceptibilities are enhanced up to 20\% in zero
field.\cite{Marcin2009}
%They found that one can have enhancement for large magnetic fields $h >
%\Delta$, but this exceeds the Pauli Hc2.\cite{Marcin2009}  
%Orientation of FFLO planes was assumed to be normal to the applied magnetic field, 
%and independent of the underlying OP nodal structure, as compared with results\cite{Vorontsov2005fflo}. 
The antiferromagnetic vector $\vq$ was found mostly to point along the FFLO
planes (i.e. $\vq \perp \vq_{FFLO}$),\cite{Yanase2009,Yanase2009abs} 
independent of whether the planes were oriented along nodes or
antinodes of the $D_{x^2-y^2}$ order parameter. 
$\vq$ across FFLO planes was not favored, except in the case of
atomic-scale {FFLO} oscillations.\cite{Marcin2009,Yanase2011}
This result might be related to the small size of the lattice grid, typically around $40\times40$ sites, 
which forces use of comparable length scales $q_{FFLO} \sim q \sim k_f$, 
but not directly applicable to superconductors with $ q_{FFLO} \sim 1/\xi_c \ll k_f$ 
(e.g. STM measurements\cite{Zhou:2013dq,Allan2013_stm115} in \cecoin\ give 
$\xi_c \sim 60 \AA$, $k_f \sim (\pi/4.6) \AA^{-1}$ and $k_f \xi_c \approx 40$).
The spatially-averaged approach\cite{Marcin2009} has also only considered
small-period modulations of the order parameter. 
In Ref.~\onlinecite{Yanase2009abs} 
appearance of AFM was correlated with presence of multiple FFLO domain walls,
but no mechanism directly linking AFM and localized ABS was established. 

The effects of the bound states have been investigated in vortex phases,
near vortex cores. The localized states in cores and 
enhancement of local density of states (LDOS) were
predicted\cite{Takigawa1999} to produce faster relaxation time $T_1$ of
electronic spins, which was later seen by spatially-resolved NMR in $S$-wave superconductor.\cite{Nakai2008} 
Bound states can result in enhancement of $T_1^{-1}$ over the normal state
value even in $D$-wave,\cite{Tanaka2015} producing 
`false Hebel-Slichter' peak below $T_c$. 
%a peak below $T_c$ of different origin than that of Hebel-Slichter. 
In Pauli-limited $D$-wave superconductors vortices can lead to SDW instability with $\vq
\parallel $nodes by increasing DOS for near-nodal directions.\cite{sdw_vortex}
Moreover, the core region of vortices often have enhanced SDW 
correlations\cite{Ogata1999,JXZhu2001,Ghosal2002} with $\vq$ across the
core, but again the role of the 
bound states for these correlations has not been explicitly shown. 

\begin{figure}
\includegraphics[scale=0.21]{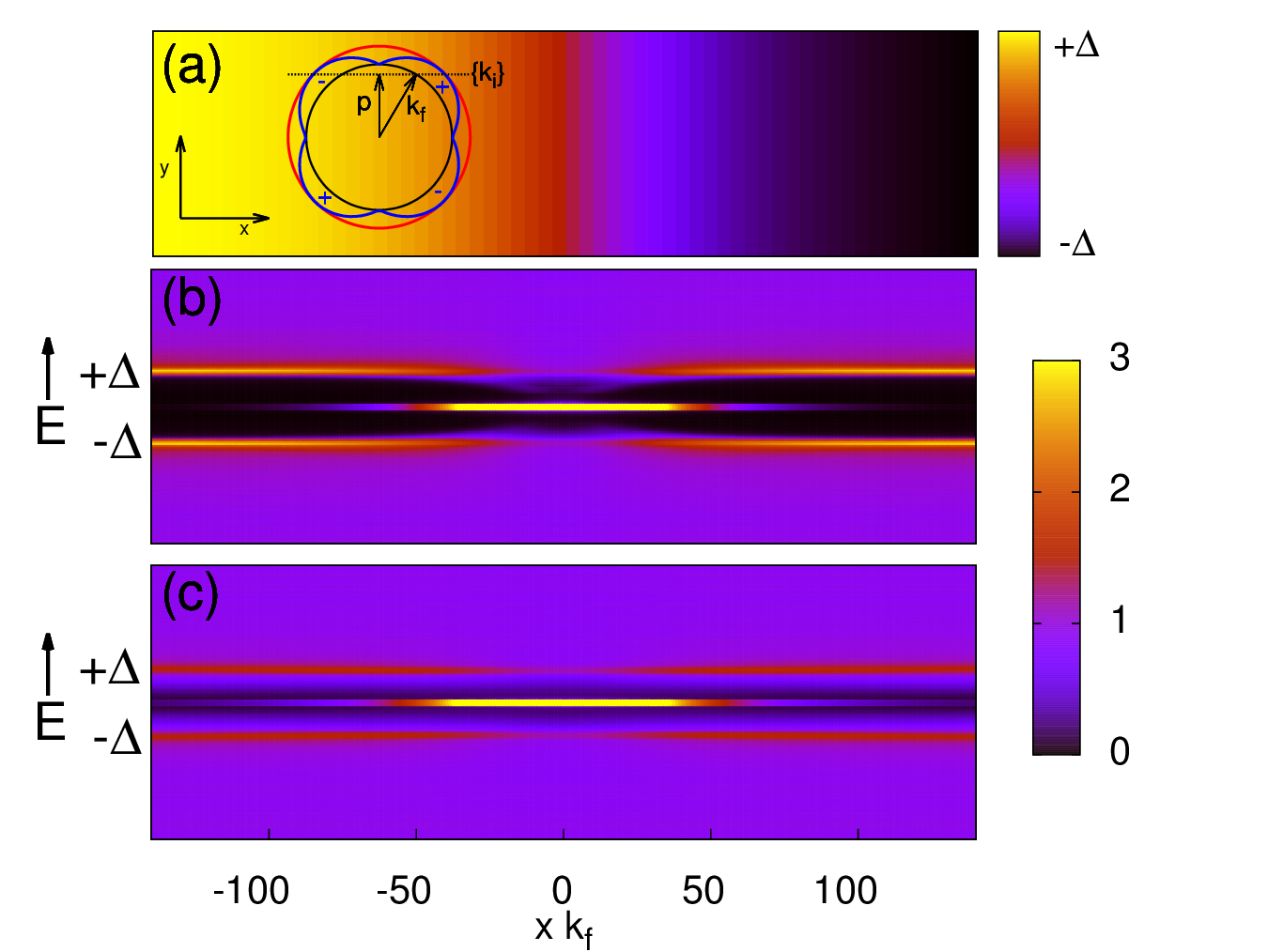}
\caption{\label{fig:1}
(a) Domain wall $+\Delta \to -\Delta$ in $x$-direction with translational
invariance along $y$. Inset shows relative orientation of the domain wall plane 
and internal symmetry of the order parameter, $S$ (red) or $D$-wave (blue). 
(b) The normalized local density of states $N(\epsilon,x)/N_f$ for $S$-wave domain wall, and  
(c) the same for $D$-wave with nodes $\parallel x$. The zero-energy states appear at the domain wall. 
We use $\Delta_0=0.05\epsilon_f$ throughout the paper.  
%Parameters are: $\Delta_0=0.05\epsilon_f$, $k_BT=\mu_B H=0.05\Delta_0$, $\omega'' = \Delta_0/100$.  
} 
\end{figure}

To clarify the role of the Andreev bound states and manifestly connect them with 
magnetic properties, we consider a prototypical non-uniform
structure of Larkin-Ovchinnikov kind: a domain wall that separates
semi-infinite regions of positive/negative amplitude of the order parameter, 
Fig.~(\ref{fig:1}).
Near the wall the density of states is
strongly peaked for zero-energy excitations, arising as result of topological
properties of Dirac-type equation.\cite{Tanaka2012,Mizushima2016} 
We consider itinerant 2-D electrons with $S$- or $D$-wave pairing symmetry. 
For $D$-wave we orient the domain wall along gap nodes,\cite{Vorontsov2005fflo} 
which also corresponds to a pairbreaking surface in a half-space problem. 
We solve the Bogoliubov-de Gennes equations in momentum space, which directly
relates the Fermi surface properties, symmetry of the order parameter, and
momentum dependence of the quasiparticle states 
to the observables. This approach also naturally connects to the quasiclassical theory. 

We find that the 
bound states lead to increase in the transverse spin susceptibility of
a superconductor 
which may lead to SDW ordering. 
The specific ordering wave vectors $\vq$ connect `hot spots' on the Fermi surface
with large bound state weights determined by coherence factors, 
that depend on the symmetry of the order parameter. 
We find that generally $S$-wave symmetry favors AFM ordering vector along the
domain wall, whereas inside $D$-wave nodally-oriented domain wall the 
ordering vector points across it. 
We also calculate relaxation rate $T_1^{-1}$ for FFLO, that so far has been lacking. 
The bound states give large relaxation rate $T_1^{-1}$ 
when quasiparticle transitions between bound states 
and continuum states can occur. 
We find that application of Zeeman field that splits the zero-energy states by $2\mu_B H$ 
generally reduces tendency toward AFM ordering inside the domain wall. 

The remainder of this report is organized as follows. In section
\ref{sec:model} we define our two-dimensional model Hamiltonian 
with a mean-field two-point order parameter $\Delta(\vx,\vx')$. 
We solve it via Bogoliubov-de Gennes equations and find quasiparticle spectrum and
amplitudes in momentum space, which we use to calculate the electron susceptibility 
and spin-lattice relaxation rate. We employ a new numeric
technique using a Fast Fourier Transform for all momenta near the Fermi surface, which is more suitable to calculating
momentum dependent quantities. 
In section \ref{sec:AandR} we present results of the calculations, and we end
in section \ref{sec:concl} with a discussion of the implications of our findings for recent experiments. 
Finally, we provide appendix \ref{app:self-cons} with outline of the self-consistent 
method we are using.

%%%%%%%%%%%%%%%%%%%%%%%%%%%%%%%%%%%%%%%%%%%%%%%%%%%%%%%%%%%%%%%%%%%%%%%%%%%%%%%%%
%%%%%%%%%%%%%%%%%%%%%%%%%%%%%%%%%%%%%%%%%%%%%%%%%%%%%%%%%%%%%%%%%%%%%%%%%%%%%%%%%

%~~~~~~~~~~~~~~~~~~~~~~~~~~~~~~~~~~~~~
\section{Model}
\label{sec:model}
%~~~~~~~~~~~~~~~~~~~~~~~~~~~~~~~~~~~~~
%
We work with the Hartree-Fock-Bogoliubov (HFB) mean-field Hamiltonian for a single band
\be
\label{eq:modelH} 
\cH_{HFB} = \frac{1}{2} \int d\vx d\vx'\quad{\Psi}^\dagger(\vx)\,{\cH}(\vx,\vx')\,{\Psi}(\vx') %+ const
\ee
where we have defined the field operator 
${\Psi}^\dag(\vx) = \left( \psi^\dag_{\uparrow}(\vx), \psi^\dag_{\downarrow}(\vx), 
\psi_{\uparrow}(\vx), \psi_{\downarrow}(\vx) \right)$, 
and ${\bf\cH}(\vx,\vx')$ is a $4\times 4$ block matrix
\be
 {\bf\cH}(\vx,\vx') = \left( \begin{array}{cc}
\hat{\cH}_{0}\,\delta(\vx-\vx') & \hat{\Delta}(\vx,\vx') \\
-\hat{\Delta}^*(\vx,\vx') & -\hat{\cH}_{0}^*\, \delta(\vx-\vx') 
\end{array} \right) \,.
\label{eq:modelHd}
\ee
$\hat{\cH}_{0}=\left[\frac{-\nabla^2}{2m^*} - \epsilon_f \right]\hat{1} - \mu_B H\sigma^z$ 
describes free electrons in a Zeeman field, 
$m^*$ is the effective mass of the electron, $\epsilon_f$ is the Fermi energy, 
$H$ is the applied magnetic field and $\mu_B$ is the Bohr magneton. 
%$\cI$ is the $2\times 2$ identity matrix, and 
$\sigma^{\alpha= \{x,y,z\}}$ are the Pauli matrices. 
The singlet superconducting pair potential is self-consistently defined as 
\bea
\hat{\Delta}(\vx,\vx') &=& (i\sigma^y)\Delta(\vx,\vx') \\
\label{eq:delta}
\Delta(\vx,\vx') &=& V(\vx-\vx')\langle \psi_{\beta}(\vx') \psi_{\alpha}(\vx) \rangle (i \sigma^y)_{\alpha\beta}
\eea
%appears in the off diagonals of ${\bf\cH}$, 
where summation over repeated spin indices is implied, and $\langle ... \rangle$ denotes ensemble average. 
$V(\vx-\vx')$ is the effective attractive interaction %between the electrons which 
that leads to superconductivity, with the cut-off energy $\Lambda$.
%and only affects electrons which are below the energy cut-off $\Lambda$.

Since we expect presense of degenerate zero-energy states we need to define 
Bogoliubov-Valatin canonical transformation with some care.\cite{MEschrig2015}
We take 
\bea
\label{eq:BdG_trans}
\left[\begin{array}{c} \psi_{\mu} \\  \psi^\dag_{\mu} \end{array} \right](\vx)
= \sum_{\vn}%{}^{'} %\left[ 
U^{(+)}_{\vn,\mu\nu}(\vx) \, \gamma_{\vn\nu} + 
U^{(-)}_{\vn,\mu\nu}(\vx) \, \gamma^\dag_{\vn\nu} 
\,, %\right] \,,
%\nonumber
\eea
where the state index $\vn$ for inhomogeneous superconductor replaces momenum $\vk$, 
used to label states in uniform superconductor. 
To treat the zero-energy states in the same way as finite-energy states, %similarly to the other states, 
$\vn$ labels all positive energy states, and \emph{half} of zero-energy states, 
as we explain below. 
The $U^{(\pm)}_{\vn}(\vx)$ are two eigenvectors of the Hamiltonian (\ref{eq:modelHd}) 
corresponding to positive and negative energy branches 
\be
\int d\vx' \cH(\vx,\vx') U^{(\pm)}_{\vn}(\vx') = \pm \epsilon_\vn U^{(\pm)}_{\vn}(\vx)
\ee
Due to particle-hole symmetry of HFB Hamiltonian, for each $\vn$ there is a pair of $\pm\epsilon_\vn$ states, 
related to each other through
\bea
U^{(+)}_{\vn,\mu\nu}(\vx) = 
\left[\begin{array}{r} \delta_{\mu\nu} \, u_\vn \\  - i\sigma^y_{\mu\nu}v_{\vn} \end{array} \right] 
\quad,\quad
%\nonumber
%\\
U^{(-)}_{\vn,\mu\nu}(\vx) = 
\left[\begin{array}{r}  - i\sigma^y_{\mu\nu}v^*_{\vn} \\ \delta_{\mu\nu} \, u^*_\vn \end{array} \right] 
\nonumber
\eea
The non-zero energy states are naturally represented by $\gamma_{\vn \mu}$ and $\gamma^\dag_{\vn \mu}$ 
terms in (\ref{eq:BdG_trans}). However as a consequence of the particle-hole symmetry, 
the zero-energy states also come in pairs, and assignment of $\gamma_{0 \mu}$ or $\gamma^\dag_{0 \mu}$ 
to them is somewhat arbitrary. 
To avoid double-counting of zero-energy states in (\ref{eq:BdG_trans}), we take half of them and assign it to 
`positive' solutions ($\gamma$, $U^{(+)}$) and the other half appear as 
`negative' part ($\gamma^\dag$, $U^{(-)}$). 
%In this way the all the states are equivalent in the transformation...
To find the positive energy states we solve 
Bogoliubov-de Gennes equations, and in case of singlet superconductivity 
they are spin-independent: 
\begin{align}
\begin{split}
& \epsilon_{\vn}u_{\vn}(\vx)= \xi(-i\grad) u_{\vn}(\vx)+\int d\vx'\Delta(\vx,\vx')v_{\vn}(\vx') \\
& \epsilon_{\vn}v_{\vn}(\vx)= -\xi(-i\grad)^* v_{\vn}(\vx)+\int d\vx'\Delta^*(\vx,\vx')u_{\vn}(\vx')
\end{split}
\label{eq:BdG}
\end{align}
where $\xi(-i\grad) =  (-i\grad)^2/ 2m^* - \epsilon_f$. 
In Zeeman field The quasi-particle excitation energy is simply shifted to %from $\epsilon_\vn$ value by 
%and the energy of a quasi-particle excitation $\gamma^\dagger_{\vn\mu}$ is 
$\epsilon_{\vn\mu} = \epsilon_{\vn} - \mu_B H \, \sigma^z_{\mu\mu}$
and the full Hamiltonaian in diagonal form is 
$\cH_{HFB} = E_0 + \sum_{\vn \mu} \; %\limits_\vn \; %_{\vn;\,\mu=\uparrow,\downarrow} \; 
\epsilon_{\vn \mu} {\gamma}_{\vn \mu}^\dag {\gamma}_{\vn\mu}$. 
%($\hat{\epsilon}_{\vn} = diag[\epsilon_{\vn\uparrow},\epsilon_{\vn\downarrow},
%-\epsilon_{\vn\uparrow},-\epsilon_{\vn\downarrow}]$)
%${\bf\gamma}_{\vn}^\dag = (\gamma^\dag_{\vn\uparrow}, \gamma^\dag_{\vn\downarrow}, 
%\gamma_{\vn\uparrow}, \gamma_{\vn\downarrow})$ 
Finally, orthogonality of solutions with $\vn\ne\vn'$, and orthogonality of 
positive and negative solutions for the same $\vn$ result in two normalization conditions: 
\bea
\int d\vx\, \left[ u_{\vn}(\vx)u_{\vn'}^*(\vx) + v_{\vn}(\vx)v_{\vn'}^*(\vx) \right] &=& \delta_{\vn\vn'} \\
\int d\vx\, \left[u_{\vn}(\vx)v_{\vn'}(\vx) - v_{\vn}(\vx)u_{\vn'}(\vx) \right] &=& 0
\eea

For the domain wall, or stripes configuration, one has translational invariance along the wall ($\hy$) 
with momentum quantum numbers $\{p\}$. 
In the transverse direction the wave function for given $p$ is expanded into Fourier Series
\be
\label{eq:wave_exp}
\begin{split}
u_{\vn}(\vx) = e^{ipy}\sum\limits_{j=0}^{N-1} \tilde{u}_{\vn}(k_j) e^{ik_jx}  \,,
\\
v_{\vn}(\vx) = e^{ipy}\sum\limits_{j=0}^{N-1} \tilde{v}_{\vn}(k_j) e^{ik_jx} \,.
\end{split}
\ee
We employ a Fast Fourier Transform technique with 
\[  \frac{k_j}{k_f} = \left\{
\begin{array}{ll}
       4\pi j/N, & j\leq N/2 \\
      -4\pi (N-j)/N, & j>N/2
\end{array} 
\right. \]
and periodic boundary conditions at $k_f x=0$ and $k_f x=N/2$ ($k_f$ is the
Fermi momentum). The reasons for beginning with a doubled Fourier domain
$(-2\pi,2\pi]$ is because the calculation of the relative momentum spin
susceptibility will half the domain to $(-\pi,\pi]$ while doubling the spatial
domain to $(-N/2,N/2)$. We use $N=2^{12}=4096$ momentum grid points.

For efficient numerics, we restrict our set of transverse momenta $\{k_j\}$ for each $p$ to include only those whose normal excitation energy 
\be
\xi(p,k_j) = \frac{k_j^2 + p^2}{2m^*} - \epsilon_f,
\ee
is below an energy cut-off, $|\xi_{p,k_j}|\leq\Lambda$. 
All higher energy solutions to (\ref{eq:BdG}) are considered normal with $\Delta(\vx,\vx') = 0$.

Furthermore, since we are interested in low-energy superconducting quasiparticles, 
we take a separable form of the pair potential, 
described by the amplitude that depends on the center of mass coordinate $\vR$, 
and the internal symmetry profile $g(\vr)$ that depends on the relative coordinate $\vr$, 
\begin{align}
\begin{split}
\vR = \frac{\vx+\vx'}{2} \qquad & \qquad \vr = \vx - \vx' \,,
\\
\Delta(\vx, \vx') = \Delta(\vR,\vr)  = & \Delta(\vR) \, \left[ \int \frac{d\vL}{(2\pi)^2}\,  g_{\hat{L}} e^{i\vL\cdot\vr} \right] 
\,,
\end{split}
\end{align}
where $\vL$ is the relative momentum in a Cooper pair. 
We consider $S$-wave and $D$-wave pairing states:  
\be
\label{eq:rot_sym}
\begin{split}
S-wave:\quad& g_{\hat{L}} = 1 \\
D-wave:\quad& g_{\hat{L}} = \sin(2\theta_{\hat{L}}) 
\quad\mbox{or}\quad g_{\hat{L}} = \cos(2\theta_{\hat{L}})
\end{split}
\ee
where $\theta_{\hat{L}}$ is the angle of $\vL$ measured from the x-axis. 
The profile of the order parameter across the domain wall depends only on coordinate $x$, $\Delta(\vR) = \Delta(x)$.
%and we further specify the profile
%$\Delta(\vR)\!\!\!\Rightarrow\!\!\!\Delta(R)$ only depends on the $\hx$ CoM
%($R=\vR\cdot\hx$) for striped geometry.

Using equations (\ref{eq:wave_exp}) for the amplitudes, (\ref{eq:BdG}) becomes
a matrix eigenvalue equation for $\epsilon_{\vn}$, where the $2N$ Fourier
coefficients, $\tilde{\cU}_{\vn}^T =
(\tilde{u}_{\vn}(k_0),\,\tilde{u}_{\vn}(k_1) ...\tilde{u}_{\vn}(k_{N-1}),
\tilde{v}_{\vn}(k_0),\,\tilde{v}_{\vn}(k_1)...\tilde{v}_{\vn}(k_{N-1}))$ 
form the eigenvector for each longitudinal momentum $p$, 
\be
\label{eq:BdG_mat}
\epsilon_{\vn}\tilde{\cU}_{\vn}
=\left( \begin{array}{cc}
\vvec{\xi}_{p} & \vvec{\Delta}_{p} \\
\vvec{\Delta}^*_{p} & -\vvec{\xi}_{p}  \end{array} \right)
 \tilde{\cU}_{\vn}
\ee
where $\vvec{\xi}_{p}$ and $\vvec{\Delta}_{p}$ are $N\times N$ matrices with $(i,j)th$ entries
\bea
\label{eq:NxN}
\vvec{\xi}_{p}(i,j) = \left(\frac{k_i^2 + p^2}{2m_e} - \epsilon_f \right)\delta_{ij} \\
\vvec{\Delta}_{p}(i,j) =  g_{\hat{L}_{ij}} \int dx\, \Delta(x) e^{-i(k_i-k_j)x}
\eea
and $\vL_{ij} = \frac{k_i+k_j}{2} \hx + p\hy$. 
Solving (\ref{eq:BdG_mat}) we obtain $2N$ eigenstates, out of which $N$ have positive (and zero) energies, 
and $N$ has mirror negative (and zero) energies. 
We arrange solutions from negative to positive 
energies, and the quantum number $\vn = (p,n)$ labels top $N$ energy states. 
This guarantees that it goes over all positive $\epsilon_\vn$ and half of zero-energy solutions. 

We consider a system where we apply a unifrom static field $\vH_0$, and consider a magnetic 
response to a small perturbation of the magnetic field 
$\delta \vH(\vx,\omega) = \int dt e^{i\omega t} \delta \vH(\vx,t)\Theta(t)$, 
where $\Theta(t)$ is the Heaviside step function. Up to first order in perturbation the electron magnetization is
\bea
\label{eq:delM}
%\vM_\alpha(\vx,\omega) = \langle \vS_\alpha(\vx,\omega)\rangle + \delta\vM_\alpha(\vx,\omega) \\
M_\alpha(\vx,\omega) = M_{0,\alpha} + \delta M_\alpha(\vx,\omega) \\
\delta M_\alpha(\vx,\omega) = \int d\vx' \; \chi_{_{\alpha\beta}}(\vx,\vx',\omega) \, \delta H_{\beta}(\vx',\omega)
\eea
where $\vM_0$ is the magnetization in the superconducting state due to the uniform field $\vH_0$. 
The bare susceptibility $\chi_{_{\alpha\beta}}(\vx,\vx',\omega)$ is 
given by the Kubo formula\cite{doi:10.1143/JPSJ.12.570}
\be 
\label{eq:sus_def}
\chi_{_{\alpha\beta}}(\vx,\vx',\omega) = i\mu_B^2\int dt \; e^{i\omega t} \, 
\langle [S_\alpha(\vx,t), S_\beta(\vx',0)]\Theta(t) \rangle 
\ee
where 
$\vS(\vx,t) = \sum_{\mu\nu} \psi_\mu^\dagger(\vx,t) \vsigma_{\mu\nu}
\psi_\nu(\vx,t)$ is the spin operator and $\omega = \omega' + i\omega''$ is
assumed to have a small imaginary part 
for convergence of the time integration 
($\omega''\ll \Delta_0$, $\Delta_0$ is the gap energy at $T=0,H=0$). 

Without effects that introduce spin-orbit coupling, the isotropy of spin space is 
broken only by $\vH_0$. 
Then the susceptibility tensor is diagonal in 
longitudinal($\delta\vH \parallel \vH_0$)-transverse($\delta\vH \perp \vH_0$) space. 
We are mostly interested in cases when the induced or spontaneous magnetization is orthogonal to 
uniform state $\delta\vM(\vq,\omega) \perp \vM_0$.
Using the Bogoliubov-Valatin transformation, the normalized transverse susceptibility is 
\be
\begin{split}
\label{eq:sus}
\chi_{_{\perp}}(\vx,\vx',\omega) = \frac{2\mu_B^2}{\chi_{_0}} \sum\limits_{\vn\vn'\mu} 
  & \left[ A_{\vn\vn'}(\vx)A^*_{\vn\vn'}(\vx') \Pi_{\vn\mu;\vn'\bmu}^{+}(\omega) \right.  \\
+ &\frac{1}{2}C^*_{\vn\vn'}(\vx)C_{\vn\vn'}(\vx') \Pi_{\vn\mu;\vn'\mu}^{-}(\omega) \\ 	
+ & \left. \frac{1}{2}C_{\vn\vn'}(\vx)C^*_{\vn\vn'}(\vx') \Pi_{\vn\mu;\vn'\mu}^{-}(-\omega) \right] 
\end{split}
\ee
Here $\bmu$ denotes spin state opposite to $\mu$, 
\be
\label{eq:fermi_factor}
\Pi_{\vn\mu;\vn'\nu}^{\pm}(\omega) = \frac{f(\epsilon_{\vn \mu}) - f(\pm\epsilon_{\vn' \nu})}
{\omega+\epsilon_{\vn \mu} \mp \epsilon_{\vn' \nu}} \,,
\ee
$f(\epsilon)$ is the Fermi distribution function, 
and $\chi_{_0}=2\mu_B^2N_f$ is the Pauli susceptibility in the normal state,  $N_f$ is the DOS
at the Fermi energy per spin projection. 
For energies close to zero, or much less than temperature spread of the Fermi-Dirac distribution, 
\be
\Pi_{\vn\mu;\vn'\nu}^{\pm}(\omega) \approx \pder{f}{\epsilon} 
\frac{\epsilon_{\vn \mu} \mp \epsilon_{\vn' \nu}}
{\omega+\epsilon_{\vn \mu} \mp \epsilon_{\vn' \nu}} 
%= \frac{1}{4T} \frac{\epsilon_{\vn \mu} \mp \epsilon_{\vn' \nu}}
= \frac{\epsilon_{\vn \mu} \mp \epsilon_{\vn' \nu}}
{4T(\omega+\epsilon_{\vn \mu} \mp \epsilon_{\vn' \nu})}  
\,.
\nonumber
\ee
Combinations of quasiparticle amplitudes 
\bea
A_{\vn\vn'}(\vx) = u^*_{\vn}(\vx)u_{\vn'}(\vx)+v^*_{\vn}(\vx)v_{\vn'}(\vx) \\
C_{\vn\vn'}(\vx) = u_{\vn}(\vx)v_{\vn'}(\vx) - v_{\vn}(\vx)u_{\vn'}(\vx)
\eea
are the coherence factors (of type II corresponding to perturbations that break time reversal symmetry\cite{tinkham}). 
They determine the spatial dependence of
susceptibility, while the remaining terms are functions of energy and
temperature. 

We note that the combinations $A_{\vn\vn'}(\vx)A^*_{\vn\vn'}(\vx')$ and
$C^*_{\vn\vn'}(\vx)C_{\vn\vn'}(\vx')$ in (\ref{eq:sus}) 
under coordinate exchange 
$\vx\leftrightarrow\vx'$ ($\vr\leftrightarrow -\vr$) become complex conjugated. 
This symmetry guarantees that local susceptibility at wave vector $\vq$ 
\be
\label{eq:sus_trans}
\chi_{}(\vR,\vq,\omega) = \int d\vr \; e^{-i\vq\cdot\vr} \chi_{}(\vR,\vr,\omega) = \chi' + i \chi''
\,,
\ee
has real part $\chi'$ that depends only on the \emph{real} part of $\Pi_{\vn\mu;\vn'\nu}^{\pm}$, 
and the imaginary part $\chi''$ has contributions only from the \emph{imaginary} part of $\Pi_{\vn\mu;\vn'\nu}^{\pm}$. 
%for a particular term $(\vn\mu;\vn'\nu)$, the contribution to 

Lastly, we find the spin-lattice relaxation rate $T_1^{-1}$ due to the
hyperfine interaction between nuclear spins $\vI(\vx_s)$ and electron spins
$\vS(\vx)$
\be
\cH_{hf} = \int d\vx d\vx_s \, \vI(\vx_s) \cdot { \cA}(\vx_s-\vx) \cdot \vS(\vx)
\ee
${ \cA}(\vr)$ is the $3\times 3$ hyperfine matrix. For transitions
between spin 1/2 nuclear states which are well below the thermal energy
($\epsilon_i-\epsilon_f=\omega<<T$), and if ${ \cA}(\vr)$ is strongly peaked near 
$\vr=0$, the spin-lattice relaxation rate due to ${\cA}_{\perp}$ is
found using first order perturbation theory, \cite{nmr_abragam}
\be
\label{eq:rel_time}
T_1^{-1}(\vR,\omega) = 2T\lim\limits_{\omega\rightarrow 0}\sum\limits_{\vq}
\quad |\cA_{\perp}(\vq)|^2 \frac{\chi''_{_{\perp}}(\vR,\vq,\omega)}{\omega}
\ee
The details of $\cA_{\perp}(\vq)$ depend on the interactions of the spin
fields, however in an effort to focus on the DW effects we consider only the
simplest isotropic coupling, $\cA_{\perp}(\vq) = A_0$. 

%~~~~~~~~~~~~~~~~~~~~~~~~~~~~~~~~~~~~~
\section{Results and Analysis}
\label{sec:AandR}
%~~~~~~~~~~~~~~~~~~~~~~~~~~~~~~~~~~~~~
We first find the profile of the order parameter for the 
domain wall configuration. The details of the self-consistent calculation are 
presented in appendix \ref{app:self-cons} and the general solution is shown in Fig.~\ref{fig:1}(a). 
The local density of states for spin projection $\mu$ is 
$ N_\mu(\epsilon,\vx) = -(1/\pi) \Im[G^R_{\mu}(\epsilon,\vx)]$ where $G^R_{\mu}(\epsilon,\vx)$ 
is the retarded Greens function, 
\begin{align} 
\begin{split}
G^R_\mu(\epsilon,\vx) = -i \int\limits_0^\infty dt\; e^{i(\epsilon+i0) t} 
\langle [\psi_{\mu}(\vx,t) \,,\, \psi_{\mu}^\dagger(\vx,0) ]_+ \rangle \\ %\Theta(t) \\ 
= \sum_{\vn}\left[ 
\frac{|u_{\vn}(\vx)|^2}{\epsilon - \epsilon_{\vn\mu} + i0} 
+ \frac{|v_{\vn}(\vx)|^2}{\epsilon + \epsilon_{\vn\bmu} +i0}\right] 
\end{split} 
%\label{eq:ldos} 
\nonumber
\end{align}
average $\langle\dots\rangle$ is over the ground state of the superconductor.  
LDOS is presented in figure \ref{fig:1}(b,c) for $S$- and $D$-wave pairings. The large zero-energy peak 
appears at the domain wall, confined on the scale of $10\xi_c$ ($\xi_c = v_f/2\pi T_c$).
In magnetic field the spectrum is Zeeman-shifted and the bound states appear at energies $\pm \mu_B H_0$ for up/down
spins. 
We perform calculations by introducing a cutoff in energy $\Lambda = 5 \Delta_0$, 
above which we treat states as if in normal metal, 
and checked that doubling of $\Lambda$ does not change our results. 
We set zero-temperature gap in terms of Fermi energy $\Delta_0 = 0.05\epsilon_f$, which results in coherence 
lengths $\xi_{c}^s = 11.2/k_f$ ($S$-wave) and  $\xi_{c}^d = 13.6/k_f$ ($D$-wave). 
The cutoff provides a rough separation of low and high energy scales, and one can break the double 
sum over $\vn$ and $\vn'$ in susceptibility (\ref{eq:sus}) into three contributions
\be
\begin{array}{c@{\qquad}l@{\qquad}l}
I & \epsilon_{\vn} <\Lambda, \; \epsilon_{\vn'}<\Lambda & \mbox{low-$\epsilon$}
\\
II  & \epsilon_{\vn} <\Lambda, \; \epsilon_{\vn'}>\Lambda; \quad (\vn\leftrightarrow \vn') & \mbox{mixed-$\epsilon$}
\\
III & \epsilon_{\vn} >\Lambda, \; \epsilon_{\vn'}>\Lambda   & \mbox{high-$\epsilon$}
\end{array}
\nonumber
\ee

%%%%%%%%%%%%%%%%%%%%%%%%%%%%%%%%%%%%%%%%%%%%%%%%%%%%%%%%%%%%%%%%%%%%%%%%%%%%%%%%%
\begin{figure*}
\includegraphics[scale=0.14]{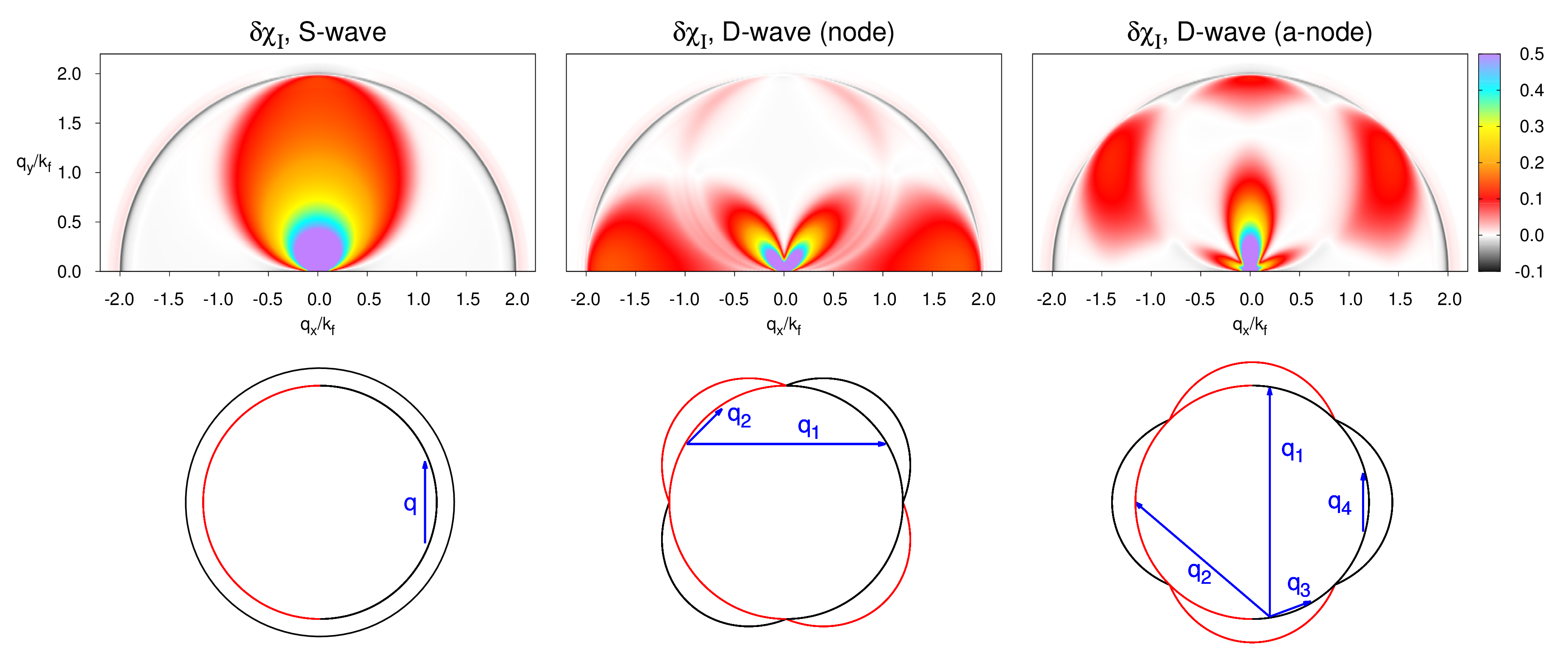}
\caption{\label{fig:2}
Upper panels show static
susceptibility $\delta\chi_{{}_I}(0,\vq,0)$ 
as a function of ordering vector $\vq$ at the center of the domain wall in the limit of low field $\mu_B
H/\Delta_0=0.01$ and temperature $k_B T/\Delta_0=0.05$. 
The purple region around $\vq=0$ (uniform magnetization) has enhancement $\delta\chi'_{{}_I}>0.5$ 
due to large density of 
bound states and has been removed to better highlight the main features. 
The $S$-wave superconductor (left) favors $\vq || \hy$, along the domain wall. 
A domain wall along nodes of $D$-wave superconductor (middle) increases tendency for AFM 
with $q_x \sim 1.75 k_f$, across the domain wall.
For antinodally-oriented domain wall (right) enhancement of $\chi$ shows for 
$\vq/k_f \sim (0,2)$ and $(1.25,1)$.
Bottom panels show ordering vectors $\vq$ that connect points on the Fermi surface with same signs of 
$\Delta_{\hk f }$ and $\Delta_{\hk' f }$ that give largest coherence factors between zero-energy bound states. 
The OP at the final end of quasiclassical trajectory $\hk$, $\Delta_{\hk f }$, 
is a product of the domain wall spatial profile (inner circle) 
and the symmetry factor $g_{\hk}$ (outer profile). black/red denote signs $\pm1$.
} 
\end{figure*}
%%%%%%%%%%%%%%%%%%%%%%%%%%%%%%%%%%%%%%%%%%%%%%%%%%%%%%%%%%%%%%%%%%%%%%%%%%%%%%%%%

%~~~~~~~~~~~~~~~~~~~~~~~~~~~~~~~~~~~~~
\subsection{Real Susceptibility}
\label{sec:realChi}
%~~~~~~~~~~~~~~~~~~~~~~~~~~~~~~~~~~~~~
We calculate the deviation of local susceptibility in non-uniform superconductor from the known normal state value 
\be
\delta\chi(x, \vq,\omega) = \chi(x, \vq, \omega) - \chi_{N}(|\vq|,\omega) \,,
\ee
which means cancellation of high-energy part $III$ in (\ref{eq:sus}). 
Mixed terms $II$ are only slightly affected by superconductivity 
and we find their contribution to $\delta\chi/\chi_N$ to be $<1\%$ for all relevant $\vq$ vectors. 
Thus, to reduce numerical cost and to obtain high-$\vq$ resolution figures, we compute 
only the dominant low-energy region terms that we denote $\delta\chi_{{}_I}$.

In figure \ref{fig:2} we show zero-field results for local static
susceptibility ($\omega=0$) in the middle of the domain wall ($x=0$) 
as a function of the ordering vector $\vq$. 
The susceptibility is clearly increased for uniform magnetization $q\approx0$,
due to large density of bound states at zero energy. There are also several
regions of non-zero $q \sim k_f$, for which $\chi_\perp$ is significantly
enhanced over the normal state value, showing tendency towards
antiferromagnetic ordering.  In $S$-wave superconductor,
Fig.~\ref{fig:2}(left), the direction of such $\vq$ vectors is along the
$y$-axis, i.e. pointing along the domain wall.

When the domain wall is along nodes of $D$-wave order parameter,
Fig.~\ref{fig:2}(middle), the ordering vector $\vq$ showing enhanced
susceptibility is along the diagonal directions for small $q_x/k_f \approx \pm
q_y/k_f$, and for $(q_x, q_y) \sim (1.75 k_f,\, 0)$ that shows about 15\% enhancement over
$\chi_0$.  The latter means that if antiferromagnetic SDW order is induced by
the non-uniform superconductivity, its modulation vector will be normal to the
order parameter domain wall, or normal to the pairbreaking surface if we
consider semi-infinite superconductor. For the domain wall in antinodal orientation, 
Fig.~\ref{fig:2}(right), enhancement appears at multiple $\vq$s, including 
$\hy$ direction similar to $S$-wave, and diagonal $\vq\sim(1.25,1)k_f$. 

We associate these regions of enhancement exclusively with correlations between bound states. 
Overall, one expects the biggest change in static $\omega=0$ susceptibility from terms in 
(\ref{eq:sus}) that have vanishing denominators of (\ref{eq:fermi_factor}) i.e.  
$\epsilon_{\vn\mu}\pm\epsilon_{\vn'\mu'} \to 0 $. 
Thus, the $(\vn\mu;\vn'\nu)$ term
which connects two bound states with zero energies should give a large contribution. 
The magnitude of this contribution, however, is also determined by the phase space, 
or the weight of zero-energy state,
and spatial dependence of the coherence factors. 
This determines the direction of $\vq$ for maximally enhanced $\delta \chi$.

To understand the role of coherence factors one can use the Andreev approximation to estimate the BdG
$u_\vn,v_\vn$ amplitudes. 
The state index can be written as $\vn=(\hk,n)$, where $\hk$ is the unit vector that defines a quasiclassical
trajectory, and $n$ labels states along this trajectory:
$$
u_\vn(\vx) = u_{\hk,n}(\vx) e^{ip_f \hk \cdot \vx}  \,,
\qquad
v_\vn(\vx) = v_{\hk,n}(\vx) e^{ip_f \hk \cdot \vx}  \,.
$$
The Andreev equations follow from BdG equations (\ref{eq:BdG}): 
\begin{align}
\begin{split}
& (\epsilon_{\hk,n} + iv_f \hk\grad) u_{\hk,n}(\vx) = \Delta(\vx,p_f\hk) v_{\hk,n}(\vx) 
\\
& (\epsilon_{\hk,n} - iv_f \hk\grad) v_{\hk,n}(\vx) = \Delta^*(\vx,p_f\hk) u_{\hk,n}(\vx) 
\end{split}
\label{eq:Andreev}
\end{align}
By approximating the domain wall profile with a step function, 
$\Delta(x, p_f \hk) = \Delta \sign(x) \; g_{\hk}$,
the amplitudes for the zero-energy bound states are, 
\be
\label{eq:uv_andreev}
\left[ \begin{array}{c}
u_{\hk,n} \\ v_{\hk,n}
\end{array} \right] (\vx)
= \frac{1}{\sqrt{2}} 
\left[ \begin{array}{c}
1  \\ -i \; \sign(\Delta^*_{\hk f}) 
\end{array} \right] 
\exp\left(- \left| \frac{\Delta g_{\hk} \,x}{ v_f \hk_x} \right| \right)
%\begin{split}
%u_{\vn}(\vx)\propto e^{ip_f \hk\cdot\vx-\kappa |\hk\cdot\vx|}e^{\frac{i}{2} [\theta_{\vn} + \phi_{\hk,\vx}]} \\
%v_{\vn}(\vx)\propto e^{ip_f \hk\cdot\vx-\kappa |\hk\cdot\vx|}e^{-\frac{i}{2} [\theta_{\vn} +\phi_{\hk,\vx}]}
%\end{split}
\ee
where $\Delta$ is the bulk amplitude of the order parameter, % in direction $\hk$, 
and $\Delta_{\hk f} = \Delta \sign(\hk_x) \, g_{\hk}$ 
is the order parameter at the \emph{final} end of the quasiclassical trajectory $\hk$. 
%$\epsilon_{\vn} = \sqrt{(v_f\kappa)^2+\Delta_{\hk}^2}$, $\theta_{\vn} =
%\sign(\hk\cdot\hx)\cos^{-1}(\epsilon_{\vn}/\Delta_{\hat{k}})$, and the energies $\epsilon_{\vn\mu} = \epsilon_{\vn} -
%\sigma^z_{\mu\mu}\mu_B H$. $\phi_{\hk,\vx}$ is the OP phase as a function of $\hat{k}$ and $\vx$.
Using this one finds that the coherence amplitudes
between bound states at points $\hk, \hk'$ on the Fermi surface 
in the middle of the domain wall ($\vx=-\vx'=\vr/2$) are 
\begin{align} 
\label{eq:AA_andreev}
\begin{split}
%A_{\vn\vn'}(\vx)A^*_{\vn\vn'}(\vx') \propto
%e^{-ip_f(\hk-\hk')\cdot\vr-2\kappa_{\vn} |\hk\cdot\vr|-2\kappa_{\vn'} |\hk'\cdot\vr|} *\cos^2(\zeta_{\vn\vn'}/2)
A^0_{\hk\hk'}(\vx) A^{0}_{\hk\hk'}(\vx')^* 
= \left|\frac{1 + \sign( \Delta_{\hk f } \Delta^*_{\hk' f } )}{2}\right|^2 \qquad 
\\
\times \; e^{-ip_f(\hk-\hk')\cdot\vr}  \; 
e^{- \frac{\Delta}{v_f} \left( \left| \frac{g_{\hk}}{\hk_x} \right| +  \left| \frac{g_{\hk'}}{\hk'_x} \right|\right) 
 | \vr\hx|}
\end{split}\end{align}
\begin{align} 
\label{eq:CC_andreev}
\begin{split}
C^0_{\hk\hk'}(\vx) C^{0}_{\hk\hk'}(\vx')^* 
= \left|\frac{1 - \sign( \Delta_{\hk f } \Delta^*_{\hk' f } )}{2}\right|^2 \qquad
\\
\times \; e^{ip_f(\hk+\hk')\cdot\vr}  \; 
e^{- \frac{\Delta}{v_f} \left( \left| \frac{g_{\hk}}{\hk_x} \right| +  \left| \frac{g_{\hk'}}{\hk'_x} \right|\right) 
 | \vr\hx|}
\end{split}\end{align}
%\be\begin{split}
%\label{eq:CC_andreev}
%C_{\vn\vn'}(\vx)C^*_{\vn\vn'}(\vx') \propto
%e^{-ip_f(\hk+\hk')\cdot\vr-2\kappa_{\vn} |\hk\cdot\vr|-2\kappa_{\vn'} |\hk'\cdot\vr|} \\ *\sin^2(\zeta_{\vn\vn'}/2)
%\end{split}
%\ee
%and $\zeta_{\vn\vn'} = \theta_{\vn}+\phi_{\hk}-\theta_{\vn'}-\phi_{\hk'}$. The
%$AA^*$ term reaches a maximum when $\zeta_{\vn\vn'} = 0$ or $2\pi$ while $CC^*$
%is maximum for $\zeta_{\vn\vn'} = \pm\pi$. Employing continuous boundary
%conditions at the DW results in the bound states having $\epsilon_{\vn} = 0$,
%$\theta_{\vn}=\sign(\hk\cdot\hx)\frac{\pi}{2}$ and we can write the $\hk$ phase
%as $\phi_{\hk} = \frac{\pi}{2}(\sign(\Delta_{\hk}) -1)$ so that
%\be
%\zeta_{\vn\vn'} = \frac{\pi}{2}[\sign(\hk\cdot\hx) +\sign(\Delta_{\hk})- \sign(\hk'\cdot\hx) - \sign(\Delta_{\hk'})]
%\ee
The ordering vector $\vq= p_f(\hk-\hk') $ that maximizes $AA^*$ in (\ref{eq:AA_andreev}) 
corresponds to combinations of $\hk$ and $\hk'$ that 
have same sign of $\Delta_{\hk f }$ and $\Delta_{\hk' f }$. 
For $CC^*$ the ordering vector is $\vq = p_f(\hk+\hk')$ and with replacement $\hk' \to -\hk'$ 
Eq.~(\ref{eq:CC_andreev}) results in the same relation between $\Delta_{\hk f }$ and $\Delta_{\hk' f }$. 
These vectors are illustrated in the bottom panel of figure \ref{fig:2}. 
For $S$-wave $g_{\hk}=1$, and the two trajectories must end up on the same side of the domain wall, 
resulting in the $\vq$ ordering generally along the domain wall. For $D$-wave, the two trajectories 
can be inside the same lobe on the same side of the domain wall giving small $q_2$ vectors, 
or there is a large wavevector $q_1 \lesssim 2 k_f$ that connects points on the mirror lobes, 
corresponding to trajectories ending up on different sides of the domain wall. 

Another slight enhancement for $D$-wave (node) can be seen as a circle of radius $k_f$ centered at $(0,k_f)$, 
especially near wavevector $\vq/k_f=(0.7,1.7)$ and the ones obtained 
by symmetry operations. 
This enhancement cannot be explained by bound states, since for these wave vectors the amplitudes 
in (\ref{eq:AA_andreev}-\ref{eq:CC_andreev}) vanish. 
We suggest that these ordering vectors correspond to 
correlations between 
the bound states and the low-energy propagating states for near-nodal directions 
$|\Delta_{\hk}| \lesssim \epsilon_{\vn} \ll \Delta_0 $. 
%The small energies are favorable to minimize the denominator in equation (\ref{eq:fermi_factor}). 
The free-propagating particle (p) and hole (h) type 
solutions $e^{\pm i k \hk\cdot \vx}$ are  
\be
\label{eq:uv_andreev_free}
\left[ \begin{array}{c}
u_{\hk,n} \\ v_{\hk,n}
\end{array} \right] %(\vx)
\propto %\frac{1}{\sqrt{2}} 
\left[ \begin{array}{c}
\epsilon + v_f k %\sqrt{\epsilon^2 - |\Delta_{\hk}|^2}   
\\ \Delta^*_{\hk} 
\end{array} \right]
e^{i k \hk\cdot \vx}
%e^{i \sqrt{\epsilon^2 - |\Delta_{\hk}|^2} \hk\cdot \vx/v_f}
\,,\,
\left[ \begin{array}{c}
\Delta_{\hk}  \\ \epsilon + v_f k %\sqrt{\epsilon^2 - |\Delta_{\hk}|^2}   
\end{array} \right]
e^{-i k \hk\cdot \vx}
%e^{-i \sqrt{\epsilon^2 - |\Delta_{\hk}|^2} \hk\cdot \vx/v_f}
\ee
with $ v_f k = \sqrt{\epsilon^2 - |\Delta_{\hk}|^2}$. 
% \be
% \label{eq:uv_andreev_free}
% \begin{split}
% u_{\vn}(\vx)\propto e^{i\vk\cdot\vx}e^{\frac{1}{2} [\pm\chi_{\vn} + i\phi_{\hk,\vx}]} \\
% v_{\vn}(\vx)\propto e^{i\vk\cdot\vx}e^{-\frac{1}{2} [\pm\chi_{\vn} + i\phi_{\hk,\vx}]}
% \end{split}
% \ee
Considering particle and hole scattering on the domain wall, we can find the exact wave functions of 
the propagating states along $\hk$. For energies near the continuum edge, the eigenvectors are 
$
\left[ \begin{array}{cc} u_{\hk,n} \,,& v_{\hk,n} \end{array} \right] 
\propto 
\left[ \begin{array}{cc} 1 \,,& \sign(\Delta_{\hk})  \end{array} \right]
$
times appropriate reflection/transmission coefficients. 
The main feature of the propagating solutions is that they are real. 
Then combination of a bound state vector (\ref{eq:uv_andreev}) for $\hk$ with 
propagating state vector for $\hk'$ results in 
\be
\label{eq:AA_bf}
A_{\hk\hk'} A_{\hk\hk'}^* \, (0,\vr) \propto 
e^{-ip_f(\hk-\hk')\cdot\vr}  \; 
e^{- \frac{\Delta}{v_f} \left( \left| \frac{g_{\hk}}{\hk_x} \right|\right) | \vr\hx|}
\ee
where dependence of the coherence factors on signs of the order parameter has disappeared, 
and similar for $CC^*$. 
Thus we have enhancement of susceptibility for vectors $\vq$ that have tails at the bottom node of the gap 
and the heads tracing the bound states along the Fermi surface. 
We note, however, that such correlations at the domain wall are weighted by the 
particle/hole transmission and reflection coefficients,  
that can be small for $\epsilon \gtrsim |\Delta_{\hk}|$. 

%%%%%%%%%%%%%%%%%%%%%%%%%%%%%%%%%%%%%%%%%%%%%%%%%%%%%%%%%%%%%%%%%%%%%%%%%%%%%%%%
\begin{figure}
\includegraphics[scale=0.1]{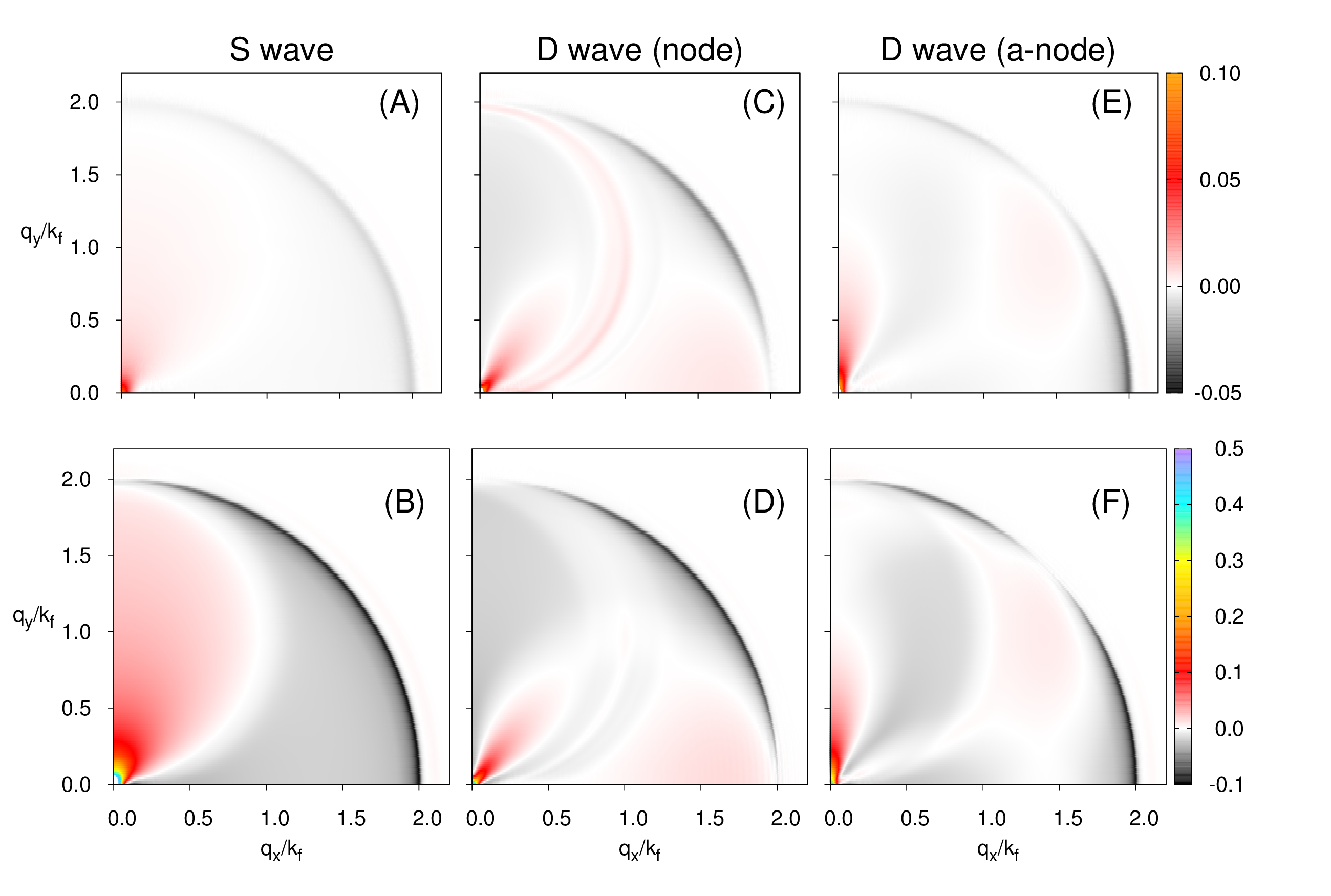}
\caption{\label{fig:sus_TB}
Effects of magnetic field and temperature on  $\delta\chi_{{}_I}(\vq)$ at the center of domain wall. 
The zero-energy peaks are shifted by $\pm\mu_BH = \pm0.4\Delta_0$, significantly reducing 
AFM correlations. 
Panels are for different temperatures: 
(A) $k_B T = 0.35\Delta_0$;  
(C,E) $k_B T = 0.2\Delta_0$; 
(B,D,F) $k_B T = 0.05\Delta_0$. 
Color scales to the right apply to the rows.
} 
\end{figure}
%%%%%%%%%%%%%%%%%%%%%%%%%%%%%%%%%%%%%%%%%%%%%%%%%%%%%%%%%%%%%%%%%%%%%%%%%%%%%%%%

In external field the energies of spin-up/down quasiparticles are shifted by $\pm \mu_B H$, 
and the zero-energy peak is split into two peaks, separated by energy $2\mu_B H$. This leads to 
reduction of $\Pi^{\pm}$ factors (\ref{eq:fermi_factor}) 
and $\delta\chi_{{}_I}$ shows very little enhancement over the normal state. 
In figure \ref{fig:sus_TB} we present $\delta\chi_{{}_I}$ at the center of domain wall for applied field
$\mu_B H = 0.4\Delta_0$, close to Pauli field, $\mu_B H_P \approx 0.7 \Delta_0\; ($S$-wave),\; 0.55\Delta_0 ($D$-wave)$. 
At lower temperatures (panels B, D and F) the zero-field enhancement regions 
are still distinguishable but are much smaller, including the $q=0$ uniform magnetization, since there is no 
zero-energy peak anymore. In $D$-wave (node) the enhancement at antiferromagnetic $q_x \sim 1.75 k_f$ is 
almost entirely wiped out. 
The higher temperature panels A, C and E
reveal a further reduction of $\chi'(\vq)$ due to a smaller self-consistent gap
and overall thermal smearing of the sum in (\ref{eq:sus}).   
We note that higher fields and temperatures mostly reduce correlations involving bound states. 
This suppression of $\delta\chi_{{}_I}$ with magnetic field at the domain wall is in stark contrast 
to behavior of susceptibility in the bulk, where magnetic field facilitates appearance of SDW 
correlations.\cite{Ikeda:2010eo,sc_afm_kato,Rosemeyer2014} 

%%%%%%%%%%%%%%%%%%%%%%%%%%%%%%%%%%%%%%%%%%%%%%%%%%%%%%%%%%%%%%%%%%%%%%%%%%%%%%%%
\begin{figure*}
\subfloat[$S$-wave \label{fig:T1_S}]{
\includegraphics[scale=0.2]{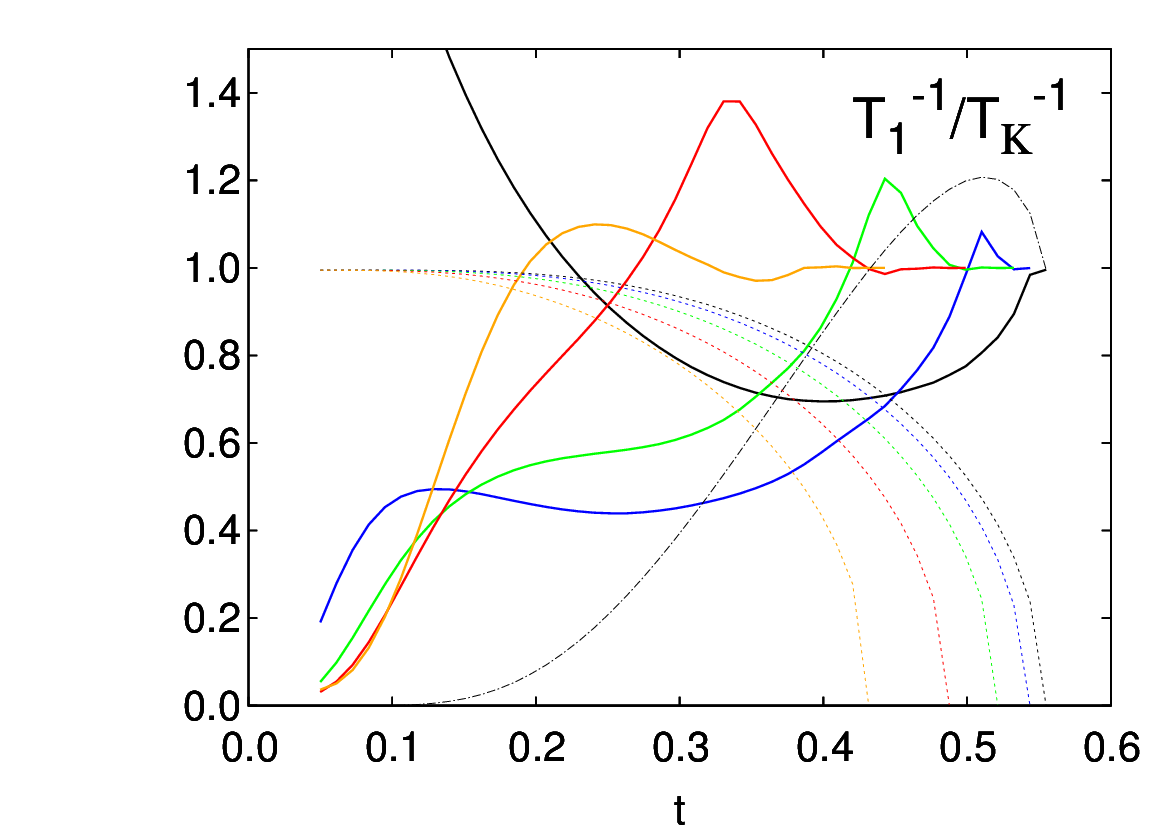} 
}
\hfill
\subfloat[$D$-wave \label{fig:T1_D}]{
\includegraphics[scale=0.2]{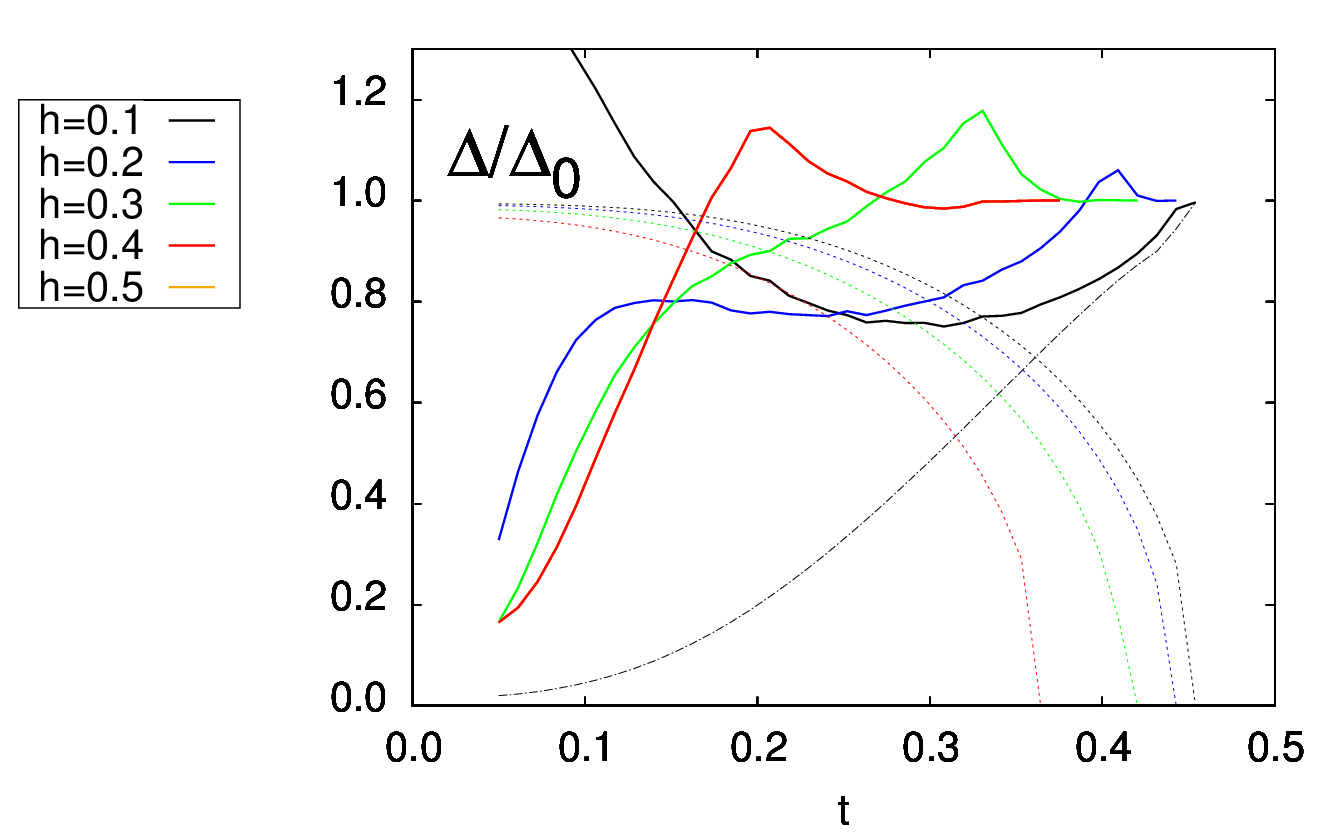} 
}
\caption{
The relaxation rate at the center of domain wall, 
normalized to the Korringa limit, $T_1^{-1} / T_K^{-1}$ (solid lines), 
and the bulk gap $\Delta/\Delta_0$ (dotted lines) as a function of temperature $t = k_B T/\Delta_0$ 
for different applied fields $h = \mu_B H/\Delta_0$. 
For higher fields, the enhancement of the relaxation rate above normal state value is due to transitions between 
bound states and the continuum states, when $\Delta(T,H) = 2 \mu_B H$ (see Fig. \ref{fig:rel_trans}), 
while at low fields the enhancement is due to transitions between bound states. 
This behavior is very different from that of the bulk relaxation rate (dot-dashed lines, shown for $h=0.1$). 
In bulk $S$-wave one can see a Hebel-Schlicter peak that is suppressed for fields above $h\sim 0.15$.
}
\end{figure*}
%%%%%%%%%%%%%%%%%%%%%%%%%%%%%%%%%%%%%%%%%%%%%%%%%%%%%%%%%%%%%%%%%%%%%%%%%%%%%%%%

%%%%%%%%%%%%%%%%%%%%%%%%%%%%%%%%%%%%%%%%%%%%%%%%%%%%%%%%%%%%%%%%%%%%%%%%%%%%%%%%
\begin{figure}
\includegraphics[scale=0.25]{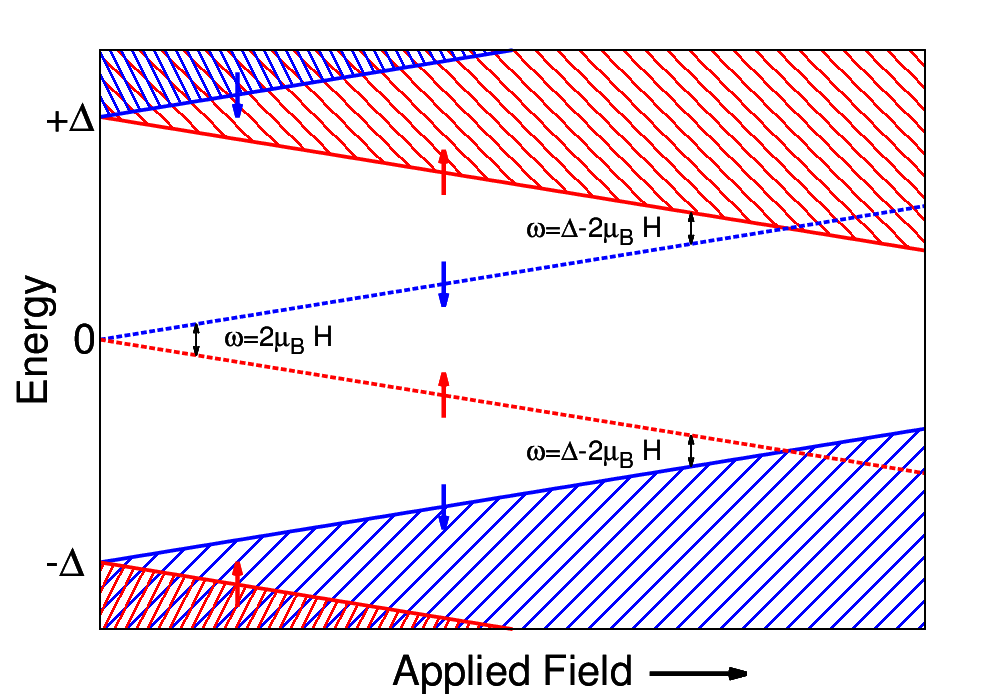} 
\caption{\label{fig:rel_trans}
Splitting of the energy states by Zeeman magnetic field. The bound states contribute to the relaxation rate 
$T_1^{-1}$ at the domain wall either at small fields, where transitions between spin-flipped bound states are allowed, 
or at fields $2\mu_B H = \Delta$ that allow transitions between bound states and the low-lying continuum 
states at $\Delta$. 
}
\end{figure}
%%%%%%%%%%%%%%%%%%%%%%%%%%%%%%%%%%%%%%%%%%%%%%%%%%%%%%%%%%%%%%%%%%%%%%%%%%%%%%%%

%~~~~~~~~~~~~~~~~~~~~~~~~~~~~~~~~~~~~~
\subsection{Relaxation Rate}
\label{sec:T1}
%~~~~~~~~~~~~~~~~~~~~~~~~~~~~~~~~~~~~~
We also calculate the imaginary part of susceptibility 
taking $\omega\rightarrow 0$ 
(well-defined for unconventional superconductors only\cite{sigrist_ueda})
\be
\begin{split}
\label{eq:sus_imag} \nonumber
&\chi_{_{\perp}}''(\vx,\vx,\omega') \propto \sum\limits_{\vn\vn'\mu} \\  %\int\,d\vr e^{-i\vq\cdot\vr} \\
&\Big[   | A_{\vn\vn'}(\vx)|^2 [f(\epsilon_{\vn\mu})-f(\epsilon_{\vn'\bmu})] 
		\delta(\omega'+\epsilon_{\vn\mu}-\epsilon_{\vn'\bmu}) %\right. 
\\
&    + \frac12 |C_{\vn\vn'}(\vx)|^2 [f(\epsilon_{\vn\mu})-f(-\epsilon_{\vn'\mu})] 
     		\delta(\omega'+\epsilon_{\vn\mu}+\epsilon_{\vn'\mu}) 
\\
& \left. - \frac12 |C_{\vn\vn'}(\vx)|^2 [f(\epsilon_{\vn\mu})-f(-\epsilon_{\vn'\mu})]
    		\delta(\epsilon_{\vn\mu}+\epsilon_{\vn'\mu}-\omega') \right]
\end{split}
\ee
to find the local spin-lattice relaxation rate (\ref{eq:rel_time}) in static limit 
$T_1^{-1}(\vR=\vx,\omega') = A_0^2 2T [\chi''_{\perp}(\vx,\vr=0,\omega')/ \omega']_{\omega'\rightarrow 0}$: 
\bea
&\displaystyle \frac{1 }{T_1(\vx) T } =  
- 2 A_0^2 \sum\limits_{\vn\vn'\mu} \pder{f(\epsilon_{\vn\mu})}{\epsilon}
\left\{ 
\left| A_{\vn\vn'}(\vx) \right|^2 \delta(\epsilon_{\vn\mu}-\epsilon_{\vn'\bmu})  
	\right.
\nonumber \\
&\hspace{2cm} 
\left. 
+ \left| C_{\vn\vn'}(\vx) \right|^2 \delta(\epsilon_{\vn\mu}+\epsilon_{\vn'\mu})
\right\} \,.
\label{eq:T1}
\eea
where for numerical evaluation we use $\delta(\epsilon) = \omega''/\pi[\epsilon^2+\omega''^2]$ 
with $\omega''=2.5\times10^{-3}\epsilon_f=\Delta_0/20$

The deviations of relaxation rate from the normal state's Korringa limit\cite{Korringa1950601} 
are due to the spin-flip transitions between the low-energy states. 
Figures \ref{fig:T1_S} and \ref{fig:T1_D}
provide numeric results for relaxation rate at the domain wall for $S$- and $D$-wave symmetry, 
with self-consistently determined bulk order parameter $\Delta(T,H)$. 
In $S$-wave one notices that the Hebel-Slichter coherence peak below $T_c$ 
for $H\to0$ is absent in the middle of the domain wall, 
due to \emph{spatial} asymmetry of the order parameter. 
However, a peak develops for higher fields, but it lies not immediately below $T_c(H)$, but at lower temperatures. 
Similar enhancement of relaxation rate above the normal state's value can also be seen in $D$-wave. 
This peak appears due to transitions between the bound states and the continuum states, when 
$\epsilon_{\vk\mu} \sim \Delta \pm \mu_B H = \mp \mu_B H = \epsilon_{0\bmu}$ 
($\omega' \to 0$ limit), as schematically shown in figure \ref{fig:rel_trans}. 

For small fields in the static limit $2\mu_B H \approx \omega' \to 0$ the 
relaxation rate is divergent due to the sharp DOS of bound states, 
that should be compared to the 
logarithmic divergence in $S$-wave bulk superconductor associated with 
the sharpness of BCS coherence peaks.\cite{tinkham}

%~~~~~~~~~~~~~~~~~~~~~~~~~~~~~~~~~~~~~
\section{Conclusions and discussion}
\label{sec:concl}
%~~~~~~~~~~~~~~~~~~~~~~~~~~~~~~~~~~~~~
%

To summarize, we found that the concentration of zero-energy Andreev bound states (in zero field) at a domain wall 
defect in the order parameter leads to significant enhancement of the bare 
susceptibility. Since variations of the order parameter occur on scale of coherence length $\xi_c \gg 1/k_f$, 
the new quasiparticle environment inside the domain wall may lead to overall divergence of the total 
\emph{local} susceptibility
$$
\chi^{RPA}(\vR,\vq) = \frac{ \chi_\perp(\vR,\vq)}{1 - J_\vq \, \chi_\perp(\vR,\vq) }
$$
for antiferromagnetic ordering vector $\vq$ ($q \sim k_f$), given sufficiently 
large exchange interaction $J_\vq$. This supports previous results of interplay between FFLO-type  
superconducting order parameter and the antiferromagnetic order in lattice models.\cite{Yanase2009abs,Marcin2009}
However, our weak-coupling approach with a single domain wall leads to results that differ 
considerably from the lattice models which used $q_{FFLO} \sim 1/\xi_c \sim q \sim k_f$. 

We find that the direction of the SDW modulation vector depends on the symmetry of 
the order parameter and the relative orientation of the domain wall and the nodes. 
For $S$-wave gap, $\vq$ is along the domain wall (i.e. $\vq \perp \vq_{FFLO}$), 
while for $D$-wave with nodes along the domain wall $\vq$-vector points across it.
The susceptibility enhancement is related to the increased correlations between 
bound states. 
These correlations disappear with magnetic field and temperature, something that was not seen in lattice models. 

Applying our results to the \cecoin\ discussion, we can say that the scenario of FFLO-induced magnetism 
is unlikely. First, the Q-phase appears in high magnetic fields\cite{cecoin5_Kenzelmann, cecoin5_Kenzelmann2} 
where we find bound state enhancement effects are wiped out. 
This high-field phase is rather more consistent with behavior of susceptibility in uniform state.\cite{sc_afm_kato,Rosemeyer2014} 
Moreover, even if the enhancement of susceptibility survives the field, 
from our calculation the direction of the SDW modulation is expected to be along the field 
(assuming $\vq_{FFLO} || nodes || \vH$), inconsistent with observations.\cite{Gerber2014}

On the other hand, in nonuniform superconductor we find an increase of the spin-lattice relaxation rate $T_1^{-1}$ 
over the Korringa limit. This enhancement mostly appears due to transitions 
between Andreev bound states and the propagating continuum states that can occur in high fields, 
$\mu_B H = 0.5 \Delta$, close to the Pauli limiting field in $D$-wave $\mu_B H_P = 0.55 \Delta_0$. 
The range of fields where it appears 
is in good agreement with experimental observations in \kbtf\ near the first-order 
superconducting-normal transition,\cite{Mayaffre2014} 
although we find the magnitude of the enhancement is somewhat smaller than the measured value. 

\section{Acknowledgements}

We thank Caroline Richard for helpful discussions and 
acknowledge support from NSF through grant DMR-0954342.

%%%%%%%%%%%%%%%%%%%%%%%%%%%%%%%%%%%%%%%%%%%%%%%%%%%%%%%%%%%%%%%%%%%%%%%%%%%%%
%%%%%%%%%%%%%%%%%%%%%%%%%%%%%%%%%%%%%%%%%%%%%%%%%%%%%%%%%%%%%%%%%%%%%%%%%%%%%
\appendix*

\section{Self-consistent order parameter}
\label{app:self-cons}

To calculate susceptibility $\chi(\vq,\vR)$, which is a function
of relative momentum $\vq$, we choose a natural momentum-based Fourier
expansion (\ref{eq:wave_exp}) to find 
self-consistent solutions of BdG amplitudes $u_\vn(\vx),v_\vn(\vx)$ from (\ref{eq:BdG}) with order parameter
(\ref{eq:delta}). 
In the past, a variety of numeric or approximate
methods have been used to address this problem: spatial lattice\cite{PhysRevB.57.8709,PhysRevLett.80.4763, Marcin2009},
Chebyshev polynomial expansion\cite{PhysRevLett.105.167006} or
quasiclassical Greens functions.\cite{Burkhardt1994, Vorontsov2005fflo} 
Though effective, they are less suitable for our purpose.

The separable order parameter $\Delta(\vx,\vx')=\Delta(\vR) g(\vr)$ 
with relative $\vr=\vx-\vx'$ and center-of-mass $\vR=(\vx+\vx')/2$ coordinates 
is obtained from mean-field definition (\ref{eq:delta}) using Bogoliubov transformation (\ref{eq:BdG_trans}):
\begin{align}
\begin{split}
\Delta(\vR) \, g(\vr) = V(\vr) \sum\limits_{\vn}{}^{'}\left\{ 
u_{\vn}(\vx) v^*_{\vn}(\vx') \left[ f(\epsilon_{\vn\downarrow}) + f(\epsilon_{\vn\uparrow}) \right] \right.
\\
\left.
- u_{\vn}(\vx') v^*_{\vn}(\vx) \left[ f(-\epsilon_{\vn\downarrow}) + f(-\epsilon_{\vn\uparrow}) \right]
\right\}
\end{split}
\label{eq:selfcon0}
\end{align}
where $f(\epsilon_{\vn\mu}) = \langle \gamma^\dagger_{\vn\mu}\gamma_{\vn\mu}
\rangle$ is the Fermi occupation number of state $\epsilon_{\vn\mu}$ with spin $\mu$. 
The prime on the sum denotes the cut-off restriction on the attractive potential $V(\vr)$, 
$|\epsilon_{\vn}|<\Lambda$,\cite{PhysRevLett.80.4763} 
which for this report we set at $\Lambda = 5\Delta_0$, 
where $\Delta_0 = 0.05\epsilon_f$ is the zero temperature bulk order parameter. 
The amplitude of the order parameter is decomposed into CoM momentum $Q$ (only $x$-component for the domain wall) 
\be
\Delta(R_x) = \int dQ \;  \tilde{\Delta}(Q) \, e^{iQR_x} \,.
\ee
Using the Fourier expanded amplitudes (\ref{eq:wave_exp}) for momenta $p$ along the domain wall, 
and $k=\{k_i\}$ in $x$ direction, and introducing relative momentum, 
$\vr \to \vq$, we write the gap equation 
\begin{align}
\begin{split}
\tilde{\Delta}(Q)\, g_{\hq} = \sum\limits_{\vn,p, k}{}^{'} \tilde{u}_{\vn}(k) \tilde{v}^*_{\vn}(k-Q) 
\qquad\qquad\qquad
\\  
\bigg\{ \tilde{V}(\vq-\vK) \left[f(\epsilon_{\vn\downarrow}) + f(\epsilon_{\vn\uparrow})\right] 
\qquad
\\
-\tilde{V}(\vq+\vK)\left[f(-\epsilon_{\vn\downarrow}) + f(-\epsilon_{\vn\uparrow})\right] \bigg\}
\end{split}
\label{eq:selfcon1}
\end{align}
Here $\vK = (k-Q/2)\hx + p\hy$, with magnitude $|\vK|,|\vq|\sim k_f$. We take separable 
interaction $\tilde{V}(\vq-\vK) = - V \, g_{\hat{q}} \, g^*_{\hat{K}}$ with a constant $V$. 
Then 
\be
\tilde{\Delta}(Q) = V \sum\limits_{\vn,p, k, \mu}{}^{'} 
\tilde{u}_{\vn}(k) \tilde{v}^*_{\vn}(k-Q)g_{\hat{K}} \tanh \left[\frac{\epsilon_{\vn\mu}}{2T}\right] 
\label{eq:selfcon2}
\ee
The interaction parameter $V$ is eliminated together with the cut-off $\Lambda$ 
using the zero temperature and field value $\Delta_0$. 
We recursively solve (\ref{eq:BdG_mat}) with 
(\ref{eq:selfcon2}) until sufficient convergence for profile  $\Delta(R_x)$ is reached. 

%%%%%%%%%%%%%%%%%%%%%%%%%%%%%%%%%%%%%%%%%%%%%%%%%%%%%%%%%%%%%%%%%%%%%%%%%%%%%
%%%%%%%%%%%%%%%%%%%%%%%%%%%%%%%%%%%%%%%%%%%%%%%%%%%%%%%%%%%%%%%%%%%%%%%%%%%%%
%~~~~~~~~~~~~~~~~~~~~~~~~~~~~~~~~~~~~~~~~~~~~~~~~~~~~~~~~~~~~~~~~~~~~~~~~~~~~~~~%
\bibliography{./mybib}
\end{document}